\documentclass{aa}  
\usepackage{txfonts}
\usepackage{hyperref} 

\usepackage{array} 
\usepackage{tabularx}

\newcommand{\dwtwo}{DE0630$-$18}
\newcommand{\dwnine}{DE0823$-$49}
\newcommand{\dwsevent}{DE1733$-$16}

\hypersetup{bookmarks=true,colorlinks=true,linkcolor=black,citecolor=black,urlcolor=blue}

\begin{document} 
   \title{Astrometric planet search around southern ultracool dwarfs}
      \subtitle{III. Discovery of a brown dwarf in a 3-year orbit around \dwtwo \thanks{\textit{Based on observations made with ESO telescopes at the La Silla Paranal Observatory under programme IDs 086.C-0680, 088.C-0679, 090.C-0786, and 092.C-0202.}}}

\author{J. Sahlmann\inst{1}\fnmsep\thanks{ESA Research Fellow}		
		\and P. F. Lazorenko\inst{2}
		\and D. S\'egransan\inst{3}
		\and E. L. Mart\'in\inst{4} 
	        \and M. Mayor\inst{3} 
		\and D. Queloz\inst{3,5} 
		\and S. Udry\inst{3}}	

\institute{European Space Agency, European Space Astronomy Centre, P.O. Box 78, Villanueva de la Ca\~nada, 28691 Madrid, Spain\\
		\email{johannes.sahlmann@esa.int}
		\and
		Main Astronomical Observatory, National Academy of Sciences of the Ukraine, Zabolotnogo 27, 03680 Kyiv, Ukraine
		\and
		Observatoire de Gen\`eve, Universit\'e de Gen\`eve, 51 Chemin Des Maillettes, 1290 Versoix, Switzerland
		\and				
		INTA-CSIC Centro de Astrobiolog\'ia, 28850 Torrej\'on de Ardoz, Madrid, Spain
		\and
		University of Cambridge, Cavendish Laboratory, J J Thomson Avenue, Cambridge, CB3 0HE, UK}

\date{Received 27 January 2015 / Accepted 26 February 2015} 

\abstract
{Using astrometric measurements obtained with the FORS2/VLT camera, we are searching for low-mass companions around 20 nearby ultracool dwarfs. With a single-measurement precision of $\sim$0.1 milli-arcseconds, our survey is sensitive to a wide range of companion masses from planetary companions to binary systems. Here, we report the discovery and orbit characterisation of a new ultracool binary at a distance of 19.5 pc from Earth that is composed of the M8.5-dwarf primary \dwtwo\ and a substellar companion. The nearly edge-on orbit is moderately eccentric ($e=0.23$) with an orbital period of 1120 d, which corresponds to a relative separation in semimajor axis of approximately 1.1 AU. We obtained a high-resolution optical spectrum with UVES/VLT and measured the system's heliocentric radial velocity. The spectrum does not exhibit  lithium absorption at 670.8 nm, indicating that the system is not extremely young. A preliminary estimate of the binary's physical parameters tells us that it is composed of a primary at the stellar-substellar limit and a massive brown-dwarf companion. \dwtwo\ is a new very low-mass binary system with a well-characterised orbit.}

\maketitle

Using astrometry with the FORS2 optical camera \citep{Appenzeller:1998lr}, installed at the Very Large Telescope (VLT) of the European Southern Observatory (ESO), we are searching for planetary companions of 20 southern ultracool dwarfs with spectral types M8--L2. The project is described in \cite{Sahlmann:2014aa}, and its first result, the discovery of a low-mass companion to an L dwarf, is reported in \cite{Sahlmann:2013ab} and updated by \cite{Sahlmann:2015_2}. Details on the astrometric reduction methods are given in \cite{Lazorenko:2014aa}.

Here, we report the discovery of the binary nature of \object{DENIS J063001.4-184014}, hereafter \dwtwo, revealed by ground-based astrometric monitoring over more than three years. The primary is a very-low-mass star or brown dwarf with an optical spectral type of M8.5 \citep{Phan-Bao:2008fr},  and the companion responsible for the astrometric orbit is a brown dwarf. 

\begin{figure}[h!]
\centering
\includegraphics[width = \linewidth]{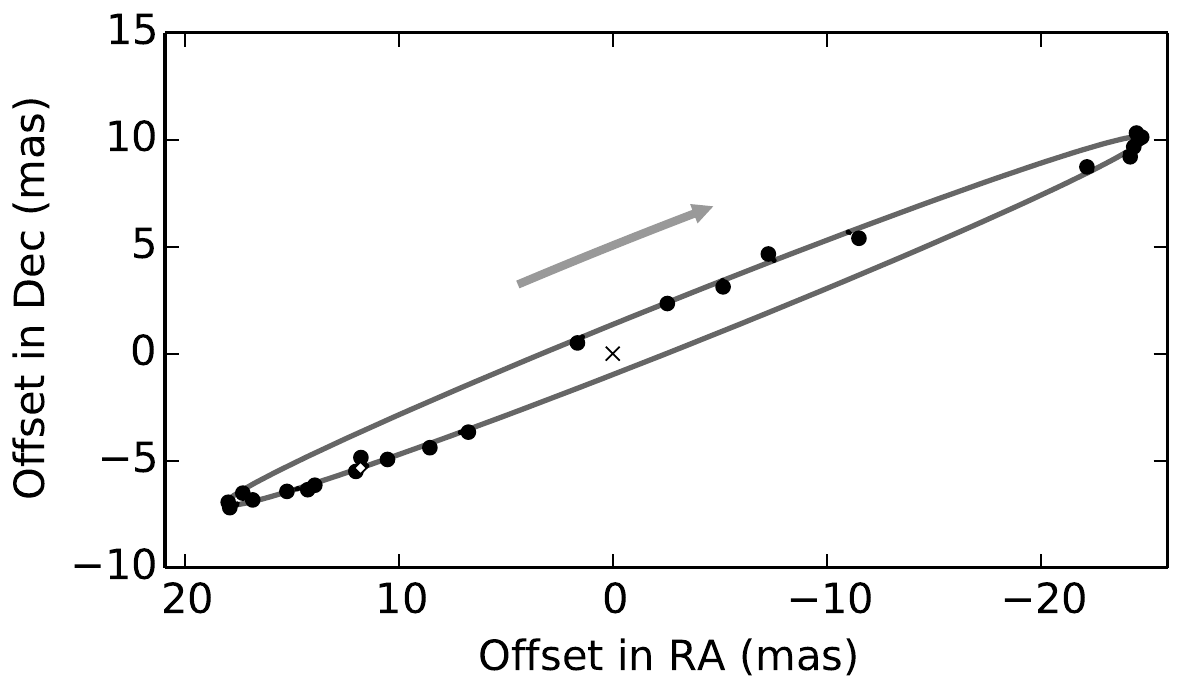}
\caption{Photocentric orbit of \dwtwo\ caused by the gravitational pull of the orbiting brown dwarf. Observations and the best-fit model are shown as black circles and a grey curve, respectively. Uncertainties are smaller than the symbol size. The barycentre and periastron position are marked with a cross and an open square, respectively. The orbital motion is clockwise; north is up and east is left.} 
\label{fig:orbit}
\end{figure}

\section{Observations and data reduction}
\dwtwo\ was observed as part of our astrometric planet search and the FORS2 data were reduced as described in \cite{Lazorenko:2014aa}. The position of the target was repeatedly measured relative to reference stars in the 4\arcmin $\times$ 4\arcmin\ field of view. \dwtwo\ has an $I$-band magnitude of 15.7 and we observed it on 23 epochs over a timespan of 1209 d between  2010 December 7 and  2014 March 30. On average, we obtained 44 frames per epoch. In June 2012, it became clear that the standard astrometric model is not sufficient to explain the motion of \dwtwo, and we initiated follow-up observations. We followed the same strategy and procedures for the adjustment of an additional Keplerian motion that is described in \cite{Sahlmann:2013ab}. A genetic algorithm was used to efficiently explore the large parameter space and to identify the most promising model parameters. These provided the starting values for a Markov-Chain Monte Carlo (MCMC) analysis that yielded the final parameters and their confidence intervals and correlation estimates.

We also observed \dwtwo\ on 2013 October 3 (MJD\footnote{Modified Julian date (MJD) is barycentric Julian date -- 2400000.5.} 56568.350793) with the red arm of UVES at the VLT \citep{Dekker:2000aa} using a $1\farcs2$ slit width, which provided a resolving power of $R$$\sim$33\,000, and the Dichroic 2 standard setup centered at 760 nm to cover the wavelength range of 565 - 931 nm. The exposure time was 2100 s and the observation took place with 0\farcs77 optical seeing at an airmass of 1.10. The spectrum was reduced using the ESO pipeline in standard setup. We measured the radial velocity on the UVES spectrum using strong atomic lines (\ion{Rb}{i} and \ion{Cs}{i}) with the method described in \citet{Sahlmann:2015_2} and determined a heliocentric radial velocity of $-13.0 \pm 1.1 $ km\,s$^{-1}$.

\begin{table}
\caption{Orbital parameters of the \dwtwo~system.}    
\label{table:1}      
\centering                       
\begin{tabular}{l c c} 
\hline\hline    
$\Delta\alpha^{\star}_0$& (mas)&    573.24$_{   -0.17}^{+    0.17}$ \\ [1pt]
$\Delta\delta_0$& (mas)&   -902.93$_{   -0.08}^{+    0.09}$ \\
$\varpi$& (mas)&     50.81$_{   -0.07}^{+    0.07}$\\
$\mu_{\alpha^{*}}$& (mas yr$^{-1}$) &    325.55$_{   -0.06}^{+    0.06}$ \\
$\mu_{\delta}$& (mas yr$^{-1}$)&   -502.75$_{   -0.04}^{+    0.03}$ \\
$e$ & $\cdots$ &      0.23$_{   -0.01}^{+    0.01}$\\
$\omega$ & (deg)&    134.33$_{   -1.35}^{+    1.42}$\\
$P$& (d)&   1120.05$_{   -2.17}^{+    2.30}$\\
$\lambda_{\rm Ref}$ & (deg)&      3.22$_{   -0.22}^{+    0.23}$\\
$\Omega$ & (deg)&    -68.32$_{   -0.12}^{+    0.12}$ \\
$i$ & (deg)&     92.73$_{   -0.15}^{+    0.15}$\\
$\alpha$& (mas)&     23.43$_{   -0.06}^{+    0.06}$ \\
$\rho$ & (mas)&     19.82$_{   -0.73}^{+    0.72}$\\
$d$& (mas)&    -26.09$_{   -0.61}^{+    0.62}$ \\
$s_\alpha$ & (mas) & {$0.186_{-0.106}^{+0.082}$}\\[1pt] 
$s_\delta$ & (mas) & {$0.157_{-0.102}^{+0.099}$}\\[3pt] 
\multicolumn{3}{c}{Derived and additional parameters}\\[3pt]
$T_\mathrm{Ref}$ &(MJD)& {56249.955966}\\[1pt]
$\Delta\varpi$& (mas)&    -0.427$_{  -0.042}^{+   0.042}$ \\
$\varpi_\mathrm{abs}$ & (mas)&     51.24$_{   -0.08}^{+    0.08}$ \\
Distance & (pc)&     19.52$_{   -0.03}^{+    0.03}$\\
\multicolumn{2}{l}{Number of epochs / frames}  & { 23 / 1005}\\
$\sigma_{O - C, \mathrm{Epoch}}$ & (mas)  & {0.239}\\ 
\hline
\end{tabular}
\tablefoot{Parameter values correspond to the median of the marginal parameter distributions and uncertainties represent 1$\sigma$ ranges. The proper motions are not absolute and were measured relative to the local reference frame.}
\end{table}

\begin{figure}
\centering
\includegraphics[width = 0.8\linewidth]{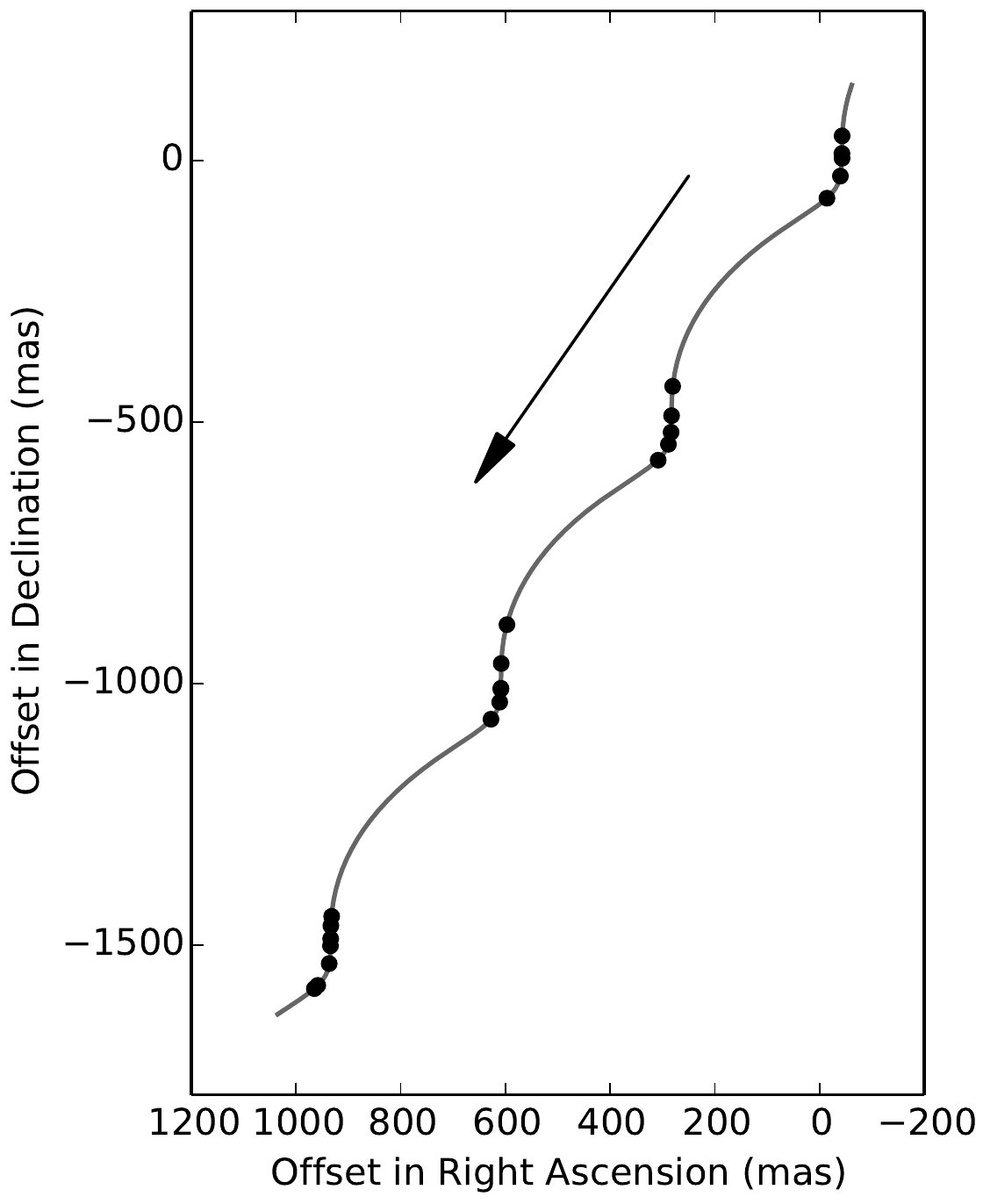}
\caption{Proper and parallactic motion of \dwtwo\ relative to the field of reference stars. The astrometric observations and the model are shown as black circles and grey curve, respectively. The black arrow indicates the direction and amplitude of the proper motion over one year.} 
\label{fig:orbitppm}
\end{figure}

\begin{figure}
\center
\includegraphics[width= 0.49\linewidth]{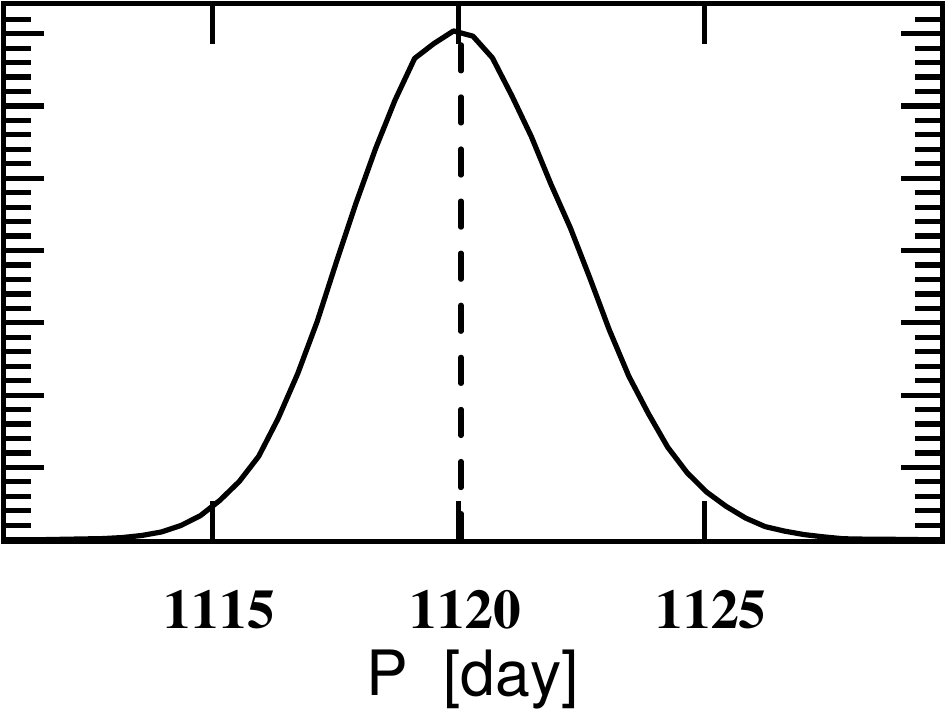}
\includegraphics[width= 0.49\linewidth]{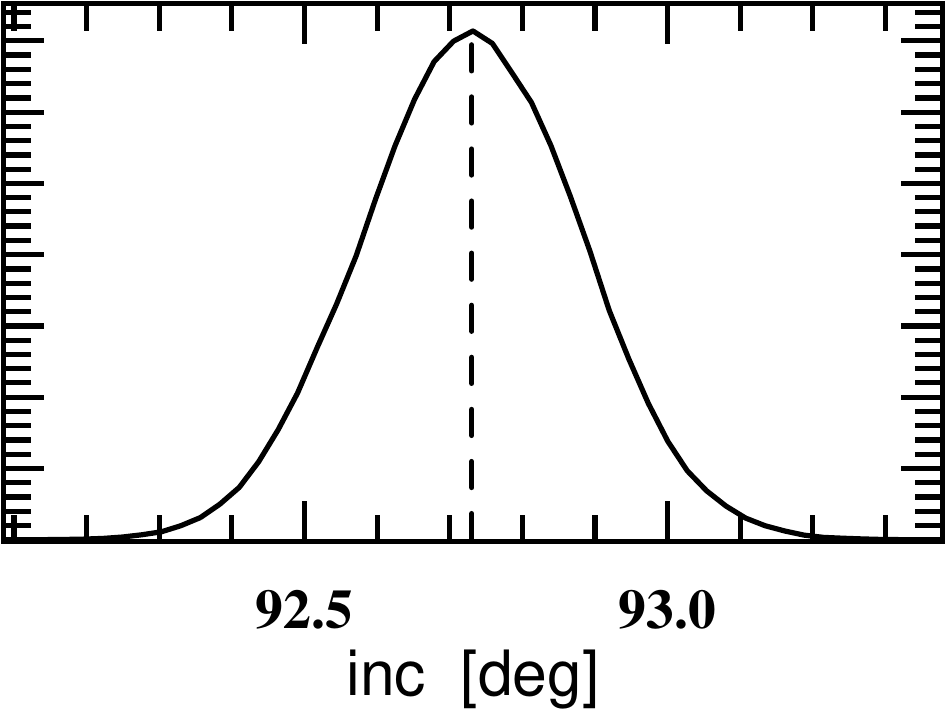}
\caption{Marginal distribution of the orbital period (\emph{left}) and inclination (\emph{right}) for \dwtwo\ obtained from $1.5 \times 10^6$ MCMC iterations. The dashed line indicates the median value and the Y-coordinate units are arbitrary and indicate relative occurrence.} 
\label{fig:marg}
\end{figure}

\begin{figure*}
\sidecaption
\includegraphics[width = 6cm]{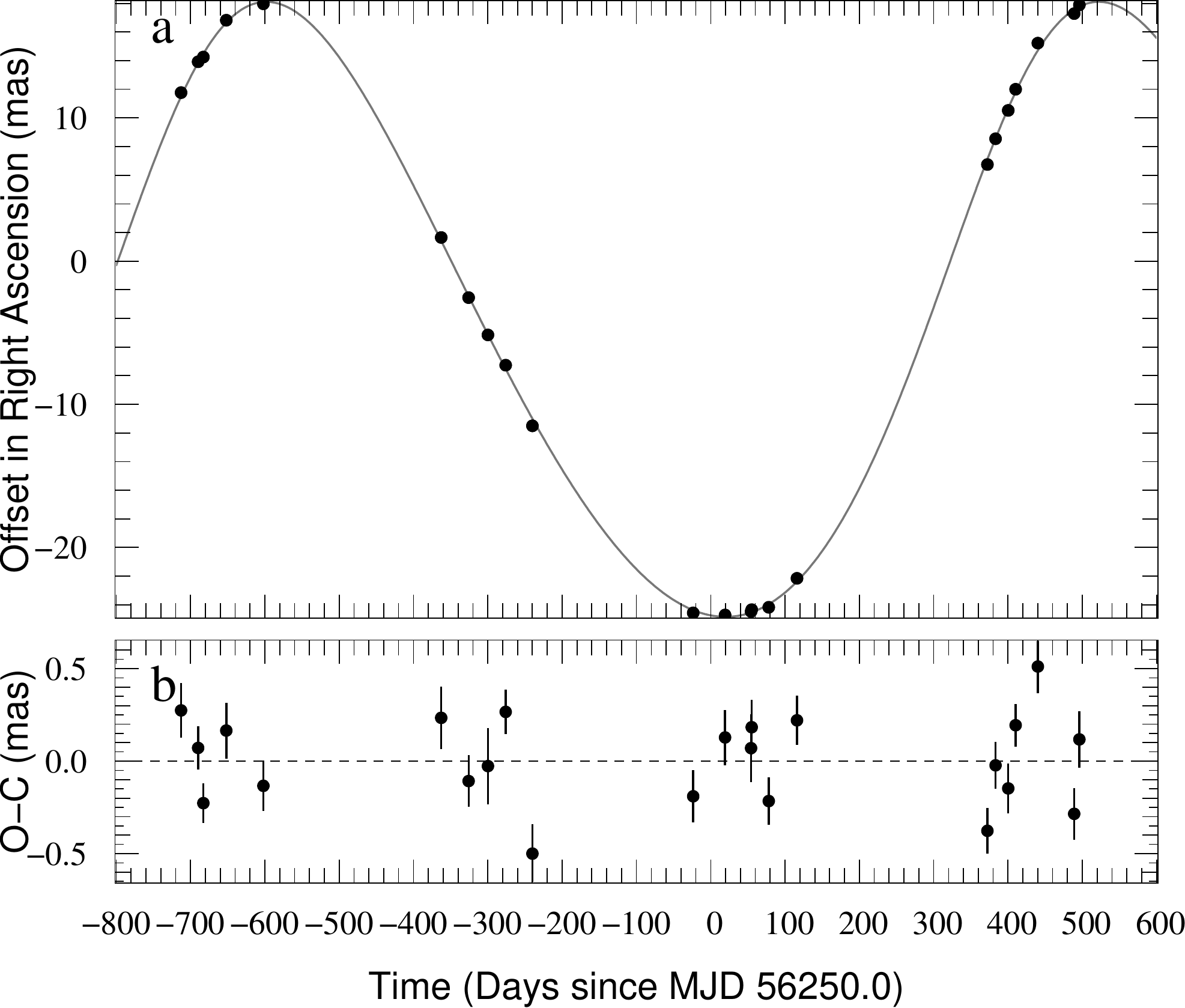}\hspace{3mm}
\includegraphics[width = 6cm]{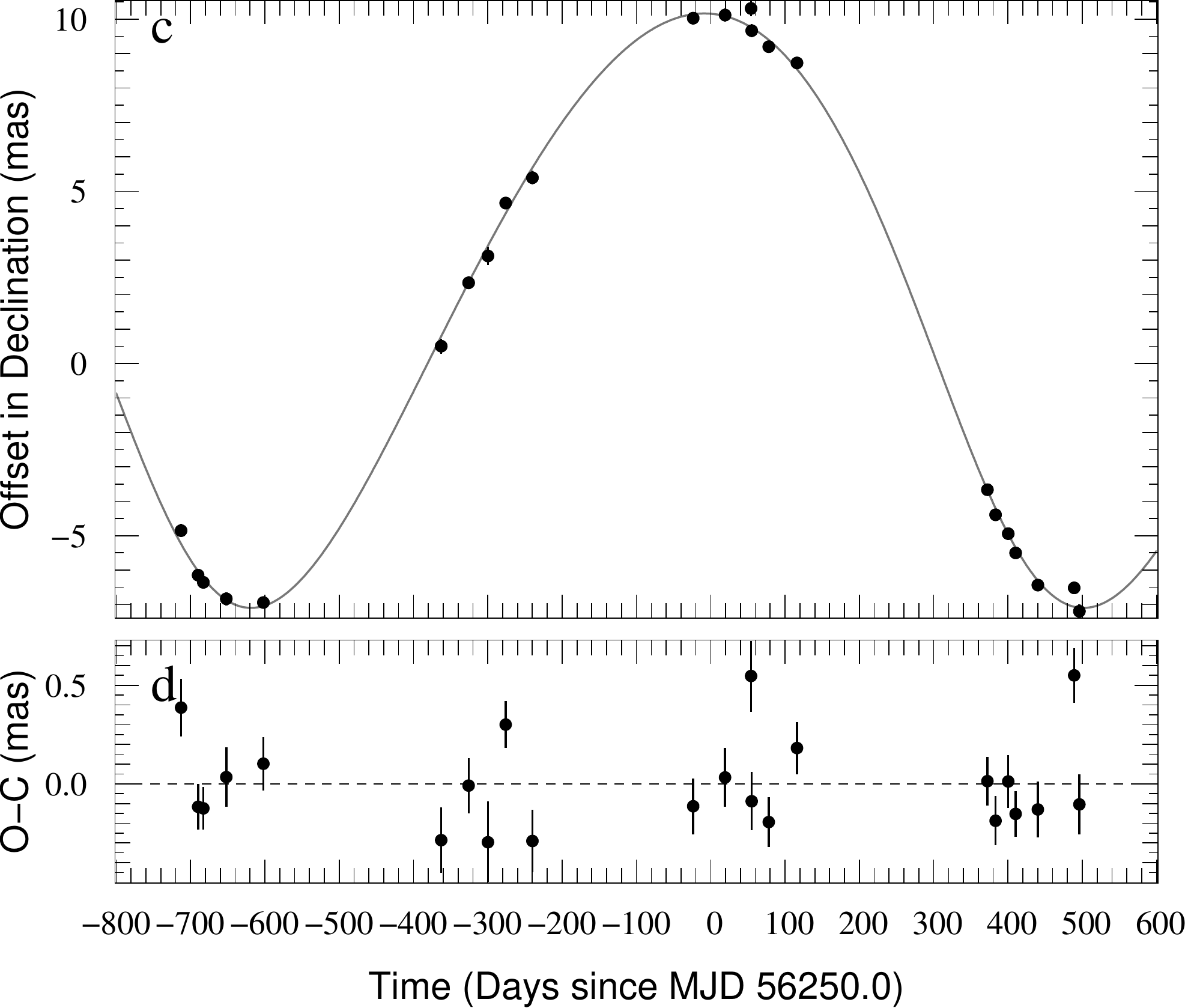}
\caption{Photocentre orbit motion of \dwtwo\ as a function of time. The orbital signature in Right Ascension (panel \textbf{a}) and declination (panel \textbf{c}) is shown, where black symbols show the epoch average values. Panels \textbf{b} and \textbf{d} show the observed minus calculated (O--C) residuals of epoch averages.}
\label{fig:axt}
\end{figure*}

\section{The photocentre orbit of \dwtwo}
The photocentre orbit of \dwtwo\ is shown in Figs. \ref{fig:orbit} and \ref{fig:axt}. Compared to the measurement precision, the photocentre semimajor axis of 23.4 milli-arcseconds (mas) is very large. The orbit is seen almost edge-on. Figure \ref{fig:orbitppm} shows the proper and parallactic motion of the \dwtwo\ system relative to the background stars, which is one order of magnitude larger than the orbital motion. 

The orbit fit parameters and their confidence intervals are reported in Table \ref{table:1}, where $\Delta\alpha^{\star}_0$ and $\Delta\delta_0$ are relative offsets to the target's position at the reference date $T_\mathrm{Ref}$ taken as the arithmetic mean of the observation dates, $\varpi$ is the relative parallax, $\mu_{\alpha^\star}$ and $\mu_\delta$ are the proper motions, $e$ is the eccentricity, $\omega$ is the argument of periastron, $P$ is the orbital period, $\lambda_{\rm Ref}$ is the mean longitude at time $T_\mathrm{Ref}$, $\Omega$ is the ascending node, $i$ is the orbit's inclination, and $\alpha$ is the semi-major axis of the photocentric orbit. The parameters $\rho$ and $d$ model the differential chromatic refraction and $s_\alpha$ and $s_\delta$ are nuisance parameters. The astrometric data constrains all orbital parameters well, which leads to normally-distributed parameters and small fractional uncertainties, see Fig. \ref{fig:marg}. The parallax correction $\Delta\varpi$ was determined in \cite{Sahlmann:2014aa} and yields the absolute parallax $\varpi_\mathrm{abs}$. The time of periastron passage $T_0$ can be retrieved via the mean anomaly 
\begin{equation}\label{eq:meanano}
M= \lambda - \omega = 2\pi \left( \frac{t}{P} -\phi_0  \right)\;\; \Rightarrow \;\;T_0 = T_\mathrm{Ref} - P \frac{M_\mathrm{Ref}}{2\pi},
\end{equation}
where t is time and $\phi_0 = T_0/P$ is the phase at periastron.

On the basis of the FORS2 astrometry, we have accurately determined the photocentre orbit of \dwtwo, in particular its semimajor axis and period. Additionally, we measured the absolute parallax, thus know the system's distance from Earth. Despite the nearly edge-on configuration, the binary is unlikely to eclipse because of the large orbital separation between two approximately Jupiter-sized bodies.

\section{Preliminary constraints on the individual components}
In \cite{Sahlmann:2014aa}, we estimated a mass of $M_1=0.086\pm0.009\,M_{\sun}$ for the primary \dwtwo A. Should the system's photocentre and barycentre coincide, the measured orbit corresponds to a mass of $M_2 \simeq0.060\,M_{\sun}$ for the companion \dwtwo B. However, at this mass ratio $q=M_2/M_1\simeq0.70$, the optical light contribution of the secondary can be significant and the photocentric and barycentric orbits may be  different.

The difference between a binary's photocentre and barycentre orbit size is determined by the magnitude difference $\Delta m$ between the components and their individual masses. The fractional mass $f = M_2 / (M_1+M_2)$ and the fractional luminosity $\beta = L_2 / (L_1+L_2) = (1+10^{0.4 \Delta m})^{-1}$, define the relationship between the semimajor axis $\alpha$ of the photocentre orbit and the semimajor axis $a_\mathrm{rel}$ of the relative orbit, where both are measured in mas: 
\begin{equation}\label{eq:3}
\alpha = a_\mathrm{rel} \, (f-\beta).
\end{equation}
An independent constraint on the relative semimajor axis is given by Kepler's law
\begin{equation}\label{eq:4}
G\, (M_1+M_2) = 4\, \pi^2 \frac{\bar a_\mathrm{rel}^3}{P^2},
\end{equation}
where $G$ is the gravitational constant, $\bar a_\mathrm{rel}$ is measured in metres and $P$ is in seconds. The relation between $\bar a_\mathrm{rel}$ and $a_\mathrm{rel}$ is given by the parallax $\varpi_\mathrm{abs}$.

Because we have knowledge of $\alpha$, $P$, and $\varpi_\mathrm{abs}$ only, the problem is underconstrained. To obtain a preliminary estimate of the possible values for individual masses, we therefore have to make use of theoretical mass--luminosity--age relationships. For a range of companion masses $M_2$ and a constant primary mass $M_1$, we used the BT-Settl \citep{Allard:2012uq} and {\small DUSTY} \citep{Chabrier:2000kx} models to obtain the corresponding magnitude difference $\Delta m_I$ in the $I$-band for ages $\geqslant$1 Gyr. Because we have measured $\alpha$, Eq. (\ref{eq:3}) yields the estimate $a^\prime_\mathrm{rel}$ for the relative semimajor axis. On the other hand, we know the orbital period and the parallax, thus Eq. (\ref{eq:4}) gives us a second estimate $a^{\prime \prime}_\mathrm{rel}$ of the relative semimajor axis. The only possible values of $M_2$ are the ones where the identity $a^{\prime \prime}_\mathrm{rel} = a^{\prime}_\mathrm{rel}$ is fulfilled.

\begin{figure}
\centering
\includegraphics[width = \linewidth]{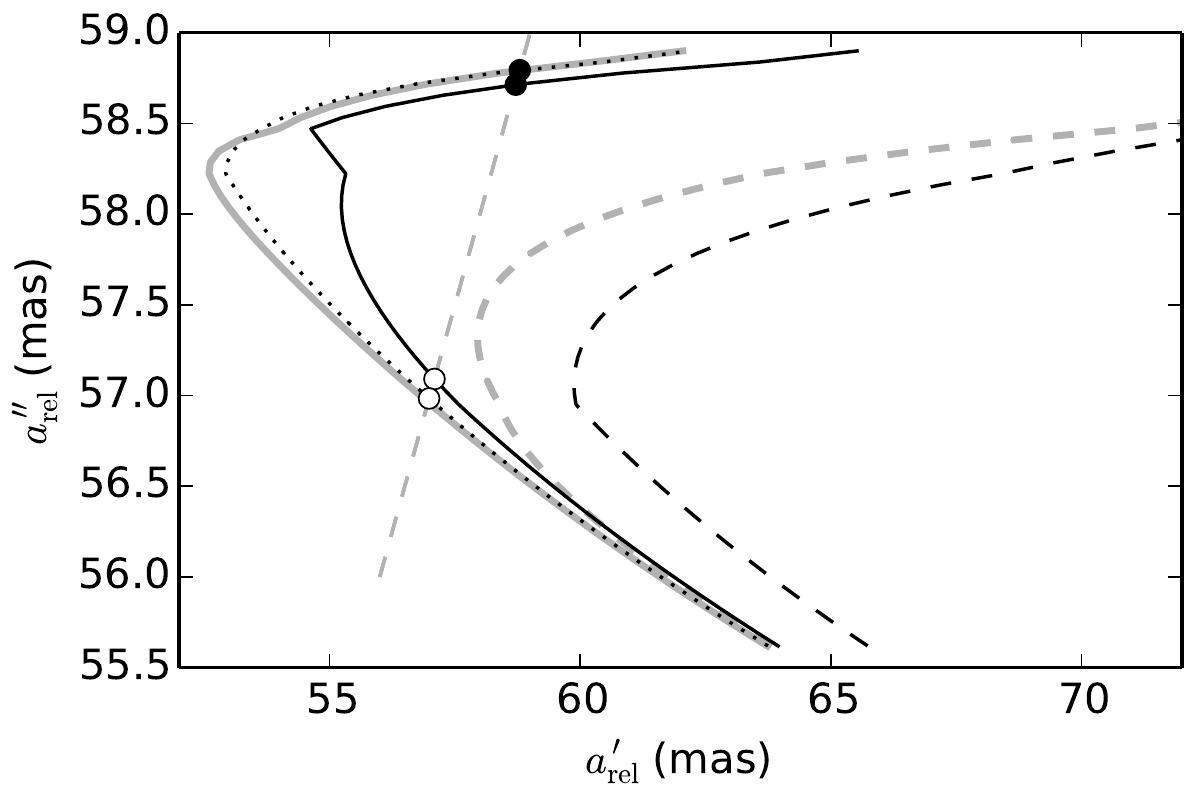}
\caption{Relative semimajor axis of \dwtwo\ obtained from two estimators using the mass--luminosity--age relationships according to BT-Settl (black curves) for ages of 1 Gyr (dashed), 3 Gyr (solid), and 7 Gyr (dotted) and DUSTY (thick grey curves) for ages of 1 Gyr (dashed) and 5 Gyr (solid). The secondary mass increases from 0.050$\,M_\sun$ to $0.076\,M_\sun$ upwards along every curve. Allowed values are marked with open and filled circles.} 
\label{fig:arel}
\end{figure}

In Fig. \ref{fig:arel}, we show the relations between $a^{\prime \prime}_\mathrm{rel}$ and $a^{\prime}_\mathrm{rel}$ for different system ages. The dashed grey line indicates equality and isolates two values of $M_2$ that are allowed for every age. The main drawback of this method is that it relies on models that may not be well calibrated in the respective mass and age range. However, it allows us to draw first preliminary conclusions on the system parameters.

According to Fig. \ref{fig:arel}, the system has to be older than 1 Gyr because the requirement $a^{\prime \prime}_\mathrm{rel} = a^{\prime}_\mathrm{rel}$  cannot be met at this age. This is compatible with the non-detection of \ion{Li}{i} absorption in our UVES spectrum of \dwtwo. Because objects with masses $\gtrsim$0.06\,$M_\sun$ deplete their lithium content within the first hundred million years of existence, the lithium test \citep[e.g.][]{Magazzu:1993kx} yields a constraint on mass and/or age of ultracool dwarfs. For \dwtwo, the Li test is negative, thus \dwtwo\ is not very young, in agreement with the constraint from the orbit modelling. 

Possible secondary masses are found at ages of 3, 5, and 7 Gyr, where the secondary can have a mass of $M_2\simeq 0.075\,M_\sun$ or of $M_2\simeq 0.060\,M_\sun$. These two solutions correspond to scenarios where the companion is very massive and luminous, thus contributes significantly to the photocentre-to-barycentre shift, and where the secondary is less massive, leading to a smaller orbit and smaller photocentre-to-barycentre shift, respectively. Table \ref{table:3} lists the model-dependent values of secondary mass, relative separation, magnitude difference in red and infrared bands, and the difference between the photocentre orbit size and the primary's barycentric orbit size ($\alpha - a_1$) for all solutions. The binary separation is in the $\sim$57--59 mas range and the magnitude differences range between 2.9 and 13.3 in the $I$-band and between 1.1 and 5.3 in the infrared $K$-band. 

\begin{table}
\caption{Theoretical constraints on secondary masses and magnitudes according to BT-Settl and DUSTY models.}
\label{table:3}      
\centering                       
\begin{tabular}{c c c c c c c}  
\hline\hline     
Age & $M_2$ & $a_\mathrm{rel}$ & $\Delta m_I$ & $\Delta m_J$   & $\Delta m_K$ & $\alpha - a_1$ \\     
(Gyr) & ($M_\sun$)& (mas)& (mag)& (mag)& (mag)& (mas)\\
\hline
\multicolumn{7}{c}{BT-Settl} \\
 3 & 0.074 &  59 & 2.9 & 1.7 & 1.2 &  -3.720 \\
 3 & 0.061 &  57 & 5.8 & 3.4 & 3.4 &  -0.279 \\
 7 & 0.075 &  59 & 2.9 & 1.6 & 1.3 &  -3.885 \\
 7 & 0.060 &  57 & 7.7 & 4.3 & 5.3 &  -0.046 \\
\hline
\multicolumn{7}{c}{DUSTY} \\
  5 & 0.075 &  59 & 2.9 & 1.7 & 1.1 &  -3.884 \\
 5 & 0.060 &  57 & 13.3 & 8.6 & 4.2 &  0.000 \\
\hline
\end{tabular}
\end{table}

\section{Discussion}
\begin{figure}[h!]
\begin{center}
\includegraphics[width=\linewidth]{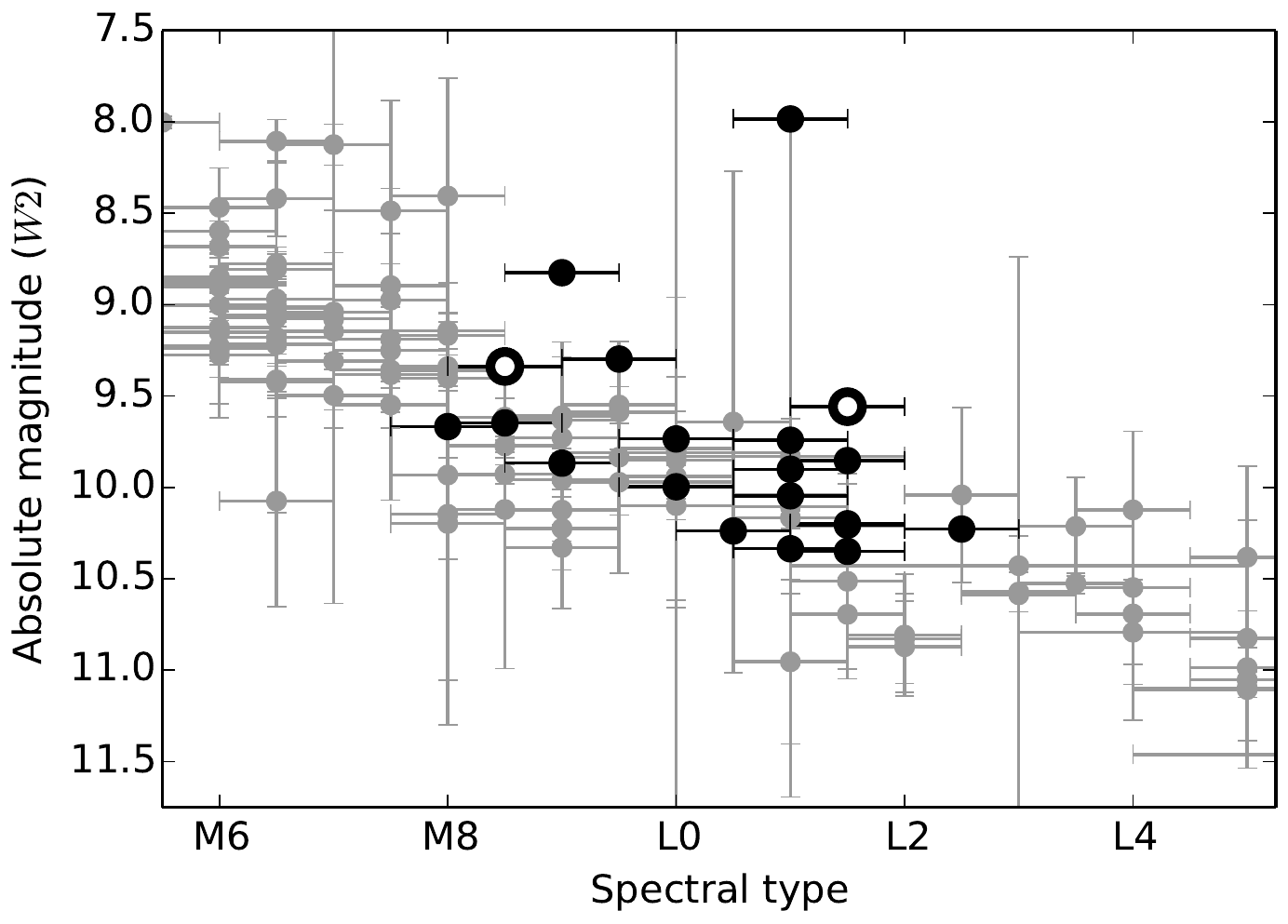}
\caption{Absolute magnitude in the $W2$-band ($\sim$$4.6\,\mu$m) as a function of spectral type for M6--L5 dwarfs in the database of ultracool parallaxes \citep{Dupuy:2012fk} (grey symbols) and for our survey sample with parallaxes from \cite{Sahlmann:2014aa} (black symbols). Magnitude uncertainties of the latter are smaller than the symbol size. Tight binaries in our sample are shown with open circles. \dwtwo\ (M8.5) is seen to have a magnitude similar to other ultracool dwarfs.}\label{fig:HRD}\end{center}
\end{figure}

Astrometric monitoring of the ultracool dwarf \dwtwo\ led to the discovery of its binary nature and allowed us to determine all orbital parameters with high precision. Because the photocentric motion in $I$-band may be diluted by the companion's emission, additional constraints are required to characterise the individual binary components. Theoretical models of substellar evolution point towards system age older than 1 Gyr and a primary mass of $M_1=0.086\pm0.009\,M_{\sun}$. The secondary mass has two allowed modes at $\sim$$0.06\,M_{\sun}$ and $\sim$$0.075\,M_{\sun}$, which in both cases indicates a substellar nature, i.e.\ the companion is a brown dwarf. From our astrometric and spectroscopic observations alone, we cannot distinguish between these two modes. However, we note that \dwtwo\ is not overluminous in the Wide-field Infrared Survey Explorer \citep[WISE;][]{Wright:2010uq} bands \textit{W1}, \textit{W2} (shown in Fig. \ref{fig:HRD}), and \textit{W3} compared to the population of ultracool dwarfs, which points towards the lower mass for the companion. In Fig. \ref{fig:HRD}, the juvenile binary \dwnine\ (L1.5, \citealt{Sahlmann:2015_2}) appears slightly overluminous. The seemingly overluminous object is \dwsevent\ (L1), which is probably caused by blending with a background source in this extremely dense starfield.

\begin{figure}[h!]
\center
\includegraphics[width= \linewidth]{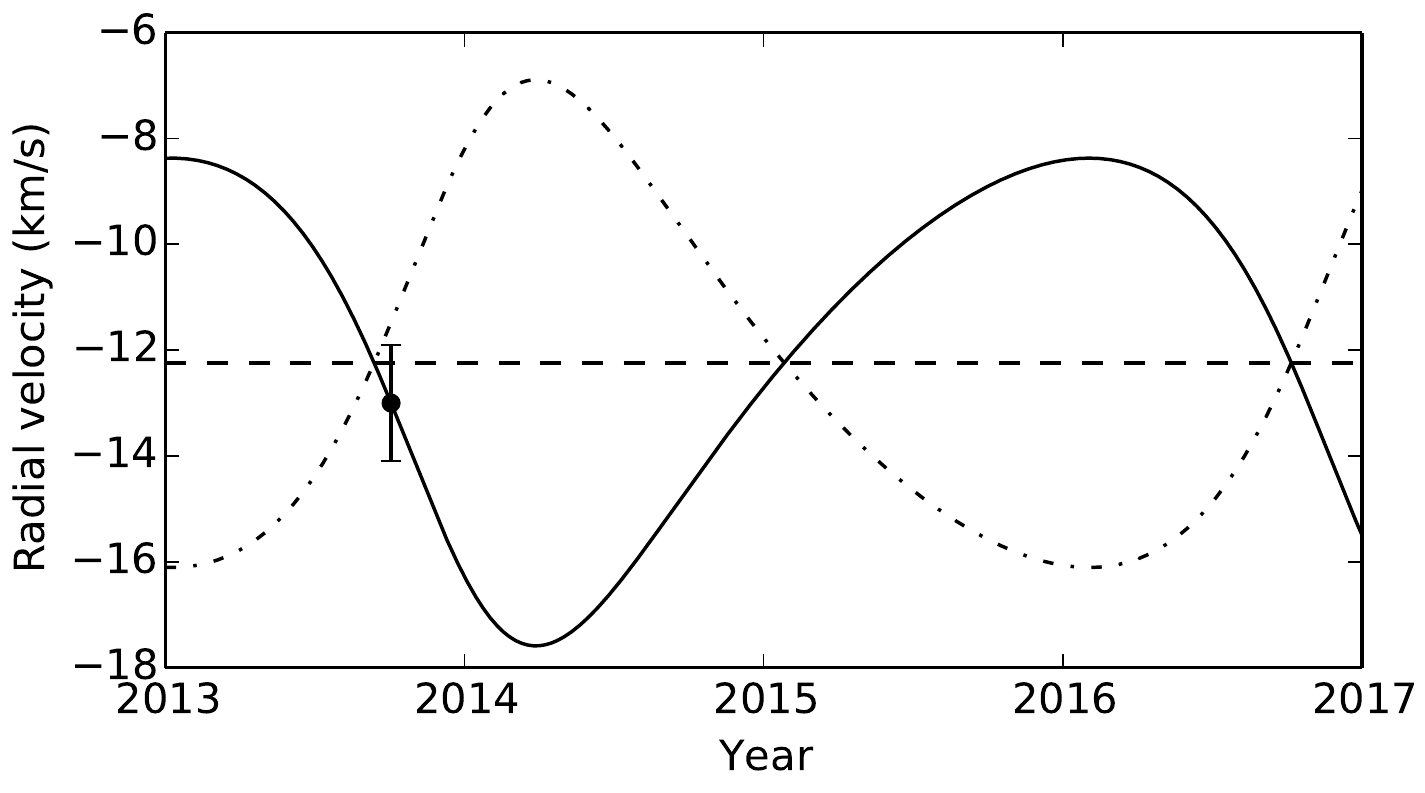}
\caption{Estimated radial velocity curve of \dwtwo A ($M_1=0.086\, M_\sun$, $M_2=0.061\, M_\sun$) as a function of time. The UVES measurement is shows with a solid circle and the systemic velocity is indicated by the horizontal dashed line. The dash-dotted line corresponds to the alternative radial velocity curve with $i^\prime = i + 180\degr$.} 
\label{fig:RV}
\end{figure}

Figure \ref{fig:RV} shows the expected radial velocity curve in this scenario. Our UVES measurement was taken close to a time of zero orbital radial velocity, which allows us to determine the heliocentric systemic velocity as either $-12.2 \pm 1.1 $ km\,s$^{-1}$ or $-13.8 \pm 1.1 $ km\,s$^{-1}$ (for the opposite orbit orientation with $i^\prime = i + 180\degr$). Clearly, the radial velocity variations along the orbit are detectable with follow-up measurements using UVES or other spectrographs.

A more detailed characterisation of the binary components of \dwtwo\ requires additional observations. The relative proximity (19.5 pc) and brightness ($m_I \simeq15.7$, $m_J \simeq11.3$) of this system will facilitate these and possibly lead to new insights into the properties of ultracool dwarfs.

The case of \dwtwo\ illustrates the potential difficulties when trying to characterise ultracool binary stars from the photocentric orbit alone, which will also arise for the hundreds of orbits \citep{Sahlmann:2014ab} expected from the astrometric survey of the \emph{Gaia} mission.

\begin{acknowledgements}
J.S. is supported by an ESA Research Fellowship in Space Science. This research made use of the databases at the Centre de Donn\'ees astronomiques de Strasbourg (\url{http://cds.u-strasbg.fr}), NASA's Astrophysics Data System Service (\url{http://adsabs.harvard.edu/abstract\_service.html}), the paper repositories at arXiv, and of Astropy, a community-developed core Python package for Astronomy \citep{Astropy-Collaboration:2013aa}. 
\end{acknowledgements}

\bibliographystyle{aa} 
\bibliography{/Users/sahlmann/astro/papers} 
\end{document}